# Observation and Structural Control of Charge-Density-Waves resonating with Terahertz Frequencies in NdNiO$_3$

Rakesh Rana,[a] Parul Pandey,[a] S.S. Prabhu[b] and D. S. Rana[a]*

[a]*Indian Institute of Science education and Research, Bhopal – 462023, India*
[b]*DCMP&MS, Tata Institute of Fundamental Research, Mumbai – 400 005, India*

Formation of charge-density-waves in ordered electronic phases (such as charge-order) is an emergent phenomenon in the perovskite class of correlated oxides. This scenario is visualized to prevail in exotic RNiO$_3$ (R=rare-earth) nickelates in which the structure controls the incipient charge-order to form in weak localization limit. However, any consequent effect demonstrating these nickelates in rare category of charge-density-waves conductors with controlled charge-lattice interactions has been a fundamental challenge so-far. Here, we present first evidence of the charge-density-waves in a prototypical NdNiO$_3$ system employing terahertz time-domain spectroscopy along selective crystal axes. A finite peak structure at 5meV in the terahertz conductivity displays all the characteristics of a charge-density-wave condensate. Contrasting charge-dynamics of collective charge-density-waves mode and Drude conductivity emerging, respectively, from orthorhombic and cubic symmetries disentangle charge-ordering from the insulating state, establish a novel structure-property cause-effect relationship, and present opportunities to harness these diverse attributes in oxide electronics.

*Email: dsrana@iiserb.ac.in

The charge-density waves (CDW) originate from the real-space modulation of electronic charge-densities to their underlying lattice.[1] These are unique source of the collective excitations upon resonant absorption and the collective charge transport induced by electric fields.[2,3] The CDW were originally found in low-dimensional systems, as attributed to the ease of formation of modulated charge densities and weak electron-phonon coupling in such systems.[1] Interestingly, these criteria are also satisfied by several high-dimensional correlated systems; however, they have rarely displayed the CDW phases. Among all, high temperature superconductors and manganites are the only systems to exhibit the formation of CDW and the impact of these limited observations has been decisive in resolving some key fundamental issues.[4-8] In superconductors, the study of CDW formed on charge-stripes of under-doped cuprates resolved the long standing issue of role electron-phonon coupling in inducing CDW vis-a-vis superconductivity.[4] In manganites, the observation of CDW and their sliding explained the magnetoresistance in a new context of field-induced collective motion of the charge-ordered phases.[2, 3,6-8] Clearly, the CDW is a much desired effect and an indispensible probe for complex electron-lattice interactions in localized electronic phases of multidimensional correlated oxides.[1-8]

In $RNiO_3$ (R = La-Lu) class of nickelates, the existence of charge-order (CO) (though debated) can be a precursor for the manifestation of the CDW.[9-20] Recently, these nickelates have attracted much attention both for their fundamental and technological fascinating attributes. For example, the interplay of charge-transfer and Mott physics presents a challenge to understand the charge/spin instabilities associated with insulator-metal (*I-M*) transition and complex magnetic-order in these systems, whereas their hetero-epitaxy induced quantum criticality holds enormous potential for emerging oxide electronics.[9-14,16,18] Among all, the $NdNiO_3$ (NNO) is a unique system as it exhibits both the *I-M* transition and the E-type antiferromagnetic order simultaneously at 205 K.[15] The Landau theory explains its magnetic structure in the framework of nested spin-density-wave system, which induces the CO as a secondary effect.[19,20] This CO has two characteristics, namely, i) its magnitude is directly proportional to the degree of orthorhombicity, and ii) it ceases to exist in tetragonal/cubic symmetries.[19-21] Experimental data too seem consistent with these propositions. For example, the observation of charge-dissproportionation, the electron-phonon coupling, two distinct time scales of the electronic relaxation, a pseudo-gap of ~ 30meV, etc, suggest a CO in orthorhombic distorted symmetry of NNO.[10,13,22,23] In contrast, no clear evidence of CO was found in coherently strained tetragonal films.[21] In a different yet relevant study, the density- and dynamic density-functional theories emphasized that large correlations in heavy rare-earth nickelates can quench the CO.[24] This, in conjunction with aforementioned Landau theory and experimental data, suggest that the CO in nickelates may well exist within weak correlations/localization limit in which case it may condense into a rare CDW mode of collective transport. This mode, if detected, will provide a fresh paradigm to the underlying physics of nickelates and can be a source of novel functionalities to the emerging electronics based on them. However, any experimental determination of this phenomenon, which unifies observation of CDW with structural control, continues to be a fundamental challenge.

In this communication, we present the first experimental evidence of the CDW as well as their structural control in classical NNO system. The CDW, when pinned moderately to lattice defects/impurities, can be resonantly excited by the electromagnetic terahertz (THz) frequencies. Hence, we have employed THz time-domain spectroscopy on epitaxial strained NNO films with different crystal structures ranging from tetragonal to orthorhombic symmetry. We show that a crossover from a CDW mode to a pure Drude-like conductivity is associated with the gradual transition from orthorhombic to tetragonal symmetry and that this CO is not necessarily a feature of the insulating state of NNO. The association of CDW with structure can be visualised as illustrated in Fig. 1 (a) and (b). The NNO epitaxial thin-films were deposited on LaAlO$_3$ (LAO) substrate with two thicknesses of 240nm and 35nm along each of *(100), (110)* and *(111)* crystal axes. These samples were abbreviated, such as a 240nm NNO *(100)* film is be referred to as *(100)*$_{240nm}$.

**Results: Structural and transport measurements:**

All the NNO films are epitaxial and phase pure (Fig. 1c). It may be noted that the LAO substrate presents an in-plane lattice mismatch ~0.3% to the bulk pseudo-cubic lattice parameter of the NNO. The strained state of the NNO films was ascertained by recording the reciprocal space maps (RSM) around asymmetric peaks (Fig. 1d and supplementary information S1). All the 35 nm films were found to be coherently strained while all the 240 nm films were in relaxed state, as shown typically for *(110)* NNO films in Fig. 1d. Furthermore, using RSM data, we determined that the *(110)* and *(111)* NNO films LAO were orthorhombically distorted with different in-plane lattice constant, while the NNO films on cubic *(100)* LAO substrate provides a tetragonal distortion with similar in-plane lattice constant for NNO [see Fig. 1 (d) and supplementary information S1].

The substrate orientation can have profound influence on stabilizing the elusive CO ground state.[25] We attempted to modulate structure and charge-ordering of NNO by growing the identical volume fraction (240nm and 35nm) in thin film form along different crystal orientations of LAO substrate. The impact of selecting dissimilar orientations of LAO on resistivity (ρ) of NNO is discernible as is summarized in Fig. 1 e and f. Though all NNO samples exhibit an *I-M* transition, their transition temperature ($T_{IM}$) is different compared to that of the bulk ($T_{IM}$~205K). This can be attributed to the effect of epitaxial strain which alters the Ni-O-Ni bond angle.[18] It is noted that the *(110)*$_{240nm}$ NNO film possesses a highest resistive state which is two and one order in magnitude larger than the *(100)*$_{240nm}$ & *(111)*$_{240nm}$ films, respectively. Also, as the NNO thickness is lowered to 35 nm, the quality of *I-M* transition varies with the crystal orientation. For instance, the *I-M* transition for *(100)*$_{35nm}$ becomes much sharper as compared to that for *(110)*$_{35nm}$ and *(111)*$_{35nm}$. Overall, electrical transport varies with both the crystal orientation and the film thickness. This presents an interesting platform to disentangle as how the effect of structure translates into the low energy dynamic response of NNO films studies employing THz spectroscopy.

**Terahertz spectroscopic investigation of low energy dynamics in NNO:**

A survey of the low energy conductivity (σ) dynamics for all the NNO films along dissimilar crystal axes and at various temperatures in the range of 5 – 320 K is presented in Fig. 2. These data depict two types of qualitative features in σ-E (E is THz photon energy) spectra, namely, a finite peak structure in one set of films and, the Drude conductivity in other set of samples. The *(110)*$_{240nm}$ and *(111)*$_{240nm}$ exhibit a pronounced peak like structure in σ-E spectra at ~5 meV and ~6 meV, respectively (Fig. 2 b and c). For *(110)*$_{35nm}$ too, a shallow peak appears at ~6meV. In contrast, the σ-E spectra of both *(100)*$_{240nm}$ and *(100)*$_{35nm}$ films, lacking peak structure is reminiscent of a Drude like conductivity (Fig. 2). A conspicuous absence of peak was noted in *(111)$_{35nm}$* film, which will be discussed later. First, we focus on the origin of peak structure in σ-E spectra of NNO *(110)* and *(111)* films. We show that this finite energy peak in the vicinity of ~5-6 meV possesses all the attributes of the collective excitations of a CDW mode. The resonance frequency of the CDW mode is in the range of ~ 1 - 8 meV with values of σ and dielectric constant (ε) of the order of $10^3$ and exhibits a typical dielectric dispersion (a zero crossing in ε at the peak σ frequency) which may be seen for *(110)$_{240nm}$* in Fig. 2.[1,7,8] These conventional CDW criteria are satisfied by the peak in σ-E spectra of *(110)$_{240nm}$*, *(111)$_{240nm}$* and *(110)$_{35nm}$* (Fig. 2e and i). However, it is imperative to discuss the possibility of optical phonon modes as they also starts emerging below the *I-M* transition of NNO.[26] In this context, the phonon vibrations of the Nd ion against the oxygen octahedral manifests at ~23 meV, a breathing phonon ~62 meV, vibrational mode ~85 meV, and the overtone intensities also do manifests but at a much higher energy range ~130 meV.[26] Thus the existence of phonon modes can be ruled out and the present observation of the peak ~5-6 meV can indeed be attributed to the CDW mode.

Another important feature is the magnitude of CDW peak for *(110)$_{240nm}$* film, which is about 3-times larger than that of the *(111)$_{240nm}$* and *(110)$_{35nm}$* film at 5K (Fig. 2a, b and d). As the CDW mode is directly linked to the strength/range of the CO correlations. Consequently, it indicates that the CO state for the *(110)$_{240nm}$* film is more robust as compared to *(111)$_{240nm}$* and *(110)$_{35nm}$* film. This is also consistent with their respective dc resistivity data of NNO films as the *(110)$_{240nm}$* film exhibits the maximum resistivity [Fig. 1 (e)]. Further, as the temperature is increased from 5K to 300K, the peak in σ gradually shifts from 4.8meV to 6.5meV for *(110)$_{240nm}$* (Fig. 2b, c and d) and from 6–7 meV for *(111)$_{240nm}$* film (Fig. 2d). This usually points towards the weakening of the CO state in the paramagnetic phase across the *I-M* transition for the NNO films (Fig. 2d).[27] Also the peak in σ shifts to higher frequencies from 5meV to 6meV as the film thickness decreases from 240nm to 35nm for the *(110)* orientation (Fig. 2b, h and k). For this reason, we speculate that the peak that appeared at ~6meV for thicker *(111)$_{240nm}$* film may have shifted to a higher frequency range >7meV for *(111)$_{35nm}$* film, which is beyond our σ-E spectra measurement range (Fig. 2k).

**Analytical treatment of the optical conductivity data: i) Drude-Lorentz model:**

The optical conductivity in σ-E spectra was modelled with phenomenological Drude-Lorentz (D-L) model (method section) for all the NNO films.[28] The experimental σ-E spectral data of *(110)$_{240nm}$*, *(111)$_{240nm}$* and *(110)$_{35nm}$* films, fitted with the D-L model at various temperatures, is shown in Fig 3(a-c). On closer examination, we find that the simulated D-L curves show a good agreement with experiment data for *(110)$_{240nm}$* and *(111)$_{240nm}$* films; however, amid a marginal deviation at low temperatures which results in a slight asymmetry of the CDW peak (Fig. 3a and 3b).[29] This asymmetry may arise due to different pinning centres present in the NNO film. Further, this deformation of the CDW mode reduces as the temperature is increased from 5K to 150K and nearly vanishes for 200K for the *(110)$_{240nm}$* and *(111)$_{240nm}$* films (shaded region Fig. 3a and 3b). Another important aspect is the CDW peak survives the *I-M* transition (~150K) of NNO, though it becomes shallower above this temperature (Fig. 3a and b). This may be attributed to the persistence of some local dynamic charge-ordering fluctuations present in the paramagnetic phase of NNO, as was also suggested by x-ray absorption spectroscopy measurements.[27] In low thickness films such as the *(110)$_{35nm}$* film, the D-L curves fit remarkably with the experimental data sans any deformation of the CDW mode (Fig. 3c). This suggests lesser number of defects pin the CDW mode in the 35nm films vis-a-vis that in thicker 240nm NNO films. As an another evidence for the CDW mode, the ε for the *(110)$_{240nm}$* film was described using, $\varepsilon(\omega) = \varepsilon_\infty + \omega_{PL}^2(\omega_{peak}^2 - \omega^2)/(\omega_{peak}^2 - \omega^2)^2 + \omega^2\Gamma^2$ where, $\omega_{PL}, \omega_{peak}, \varepsilon_\infty, \Gamma$ are the plasma frequency, peak frequency of the CDW mode, dielectric constant of the bound electron at infinity (Fig. 3d).[1,8] Further, the *(100)$_{240nm}$* and *(100)$_{35nm}$* exhibit only the Drude like conductivity. Hence, their spectrum was fitted only to the Drude part of the D-L model and is shown for *(100)$_{240nm}$* film in Fig. 3e. A summary of the D-L fitted parameters for *(110)$_{240nm}$* and Drude fits for *(100)$_{240nm}$* is given in the supplementary information Table S2.

**(ii) $T^2$ dependence of scattering rate and gap opening:**

To confirm the origin of the peak in the CDW excitations, the scattering rate ($\Gamma$) versus temperature (T) should obey the relation, $\Gamma(T) = \Gamma_o + AT^2$, where, $\Gamma_o$ is the $\Gamma(T \to 0K)$ and A is the constant [Fig 4a and inset Fig 4(a)].[1,8] We find for *(110)$_{35nm}$*, *(110)$_{240nm}$* and *(111)$_{240nm}$* films with CDW excitations, the experimental data fits well with this equation below the *I-M* transition (~ 150K) (Fig. 4a). The fitted values thus obtained are $\Gamma_o$ =3.1, 1.6, and 2.08 meV, while A = 0.69, 0.015, 0.04 meV/K$^2$ for the *(110)$_{35nm}$*, *(110)$_{240nm}$* and *(111)$_{240nm}$*, respectively. These values occur in the similar range as of the classic 3d CDW systems.[7,8] Next, we show that the CDW excitations in the NNO film are indeed associated with opening of a pseudo-gap below the single particle excitation spectrum. To elucidate this, we normalized σ at 0.8 meV and at 5meV with respect to the room temperature σ for the *(110)$_{240nm}$* (Fig. 4b). It may be seen for *(110)$_{240nm}$* film that normalized σ exhibits only a marginal change below *I-M* at 0.8meV but it increases nearly at low temperatures to three times its value at 300K at 5meV (Fig. 4b). Similar behaviour was noted for Pr$_{0.7}$Ca$_{0.3}$MnO$_3$ films.[8] This suggests that the observed CDW mode may be associated with opening of a 2Δ gap at the Fermi level (Fig. 4b).

The values of Δ can be estimated by fitting a thermally activated hopping model to the optical conductivity σ(T) and Drude plasma frequency [$\omega_{PD}(T)$] in the low frequency range (*i.e.* 0.4THz) where Drude carriers are more activated, using the equation $\sigma(T) = A_o \exp[-\Delta/k_\beta T] + B$, here $A_o$ is a constant, Δ is the activation energy, $k_\beta$ is Boltzmann constant and B is the non-zero offset (inset Fig. 4b and Fig. 4c).[30] The best fitted values of Δ after fitting $\sigma(T)$, $\omega_{PD}(T)$, for *(110)$_{35nm}$* is 24(±4) meV, *(110)$_{240nm}$* is 25±3 meV and *(111)$_{240nm}$* is 29±6 meV. These distinct values of Δ for NNO films may arise due to difference in quality of samples.[31] Nevertheless; this is a strong indication of the electron-phonon coupling. It is remarkable that the obtained values of Δ follow standard BCS picture of the gap opening with 2Δ~3.5$k_\beta T_N$.[32] Using this relation for bulk NNO, a Δ~30meV for $T_N$~205K and a Δ~22meV for a $T_N$~150K for present NNO films may be approximated [supplementary information S3]. These calculated values of Δ for NNO films are lesser than a SDW gap measured using optical spectroscopy of ~200 meV, however are in good agreement with a Δ of ~15meV (using $T_N$~100K) measured in the tunnelling conductance measurements and a Δ of 44meV found using Fourier transform infrared spectroscopy.[9,11,23] Thus our experimental observations for NNO are suggestive of a BCS theory based picturesque for the states near the Fermi level.

### iii) Temperature dependence of Dynamic THz Conductivity:

The temperature dependence of σ also provides vital clues about the underlying mechanisms driving the NNO films and for selected photon energies (2, 3, 4, 5 and 6meV) is shown in Fig. 4 d and e. The *I-M* transition for the *(100)*$_{35nm}$ is much sharper than the *(110)*$_{35nm}$ and *(111)*$_{35nm}$ films (Fig. 4d) which further displays a direct correspondence with the dc transport (ρ-T) data (Fig. 1f). Similarly, the *(100)*$_{240nm}$ sample though exhibits a broad *I-M* transition, nevertheless remains more conducting vis-à-vis *(110)*$_{240nm}$ and *(111)*$_{240nm}$ films (Fig. 4e). However, for *(110)*$_{240nm}$ film rather than observing a decrease in σ below *I-M* transition, an increase in σ at 3meV and 5meV (Fig. 4e) is assigned to the CDW mode. Similarly, for *(110)*$_{35nm}$ and *(110)*$_{240nm}$ NNO films, large values of σ at ~6meV arise due to a dominant CDW contribution for these films which is superimposed on the Drude contribution, while later is the only driving mechanism for the *(100)* oriented NNO films.

### iv) Spectral weight analysis:

The spectral weight (SW) of the Lorentz and Drude part was calculated from the experimental data (Fig. 5). The SW can effectively track the charge redistribution due to optical sum rule as per which the integral of σ over all the frequencies remains constant and can provide the valuable insights of the condensed (in the CDW phase) and uncondensed electrons (Drude carriers). The SW weight was deduced using the equation, $SW=\int_{\omega 1}^{\omega 2}\sigma(\omega)d\omega$, where $\omega_1$ =0.5meV and $\omega_2$ =7meV are the upper and lower cut off frequencies.[7] It may be seen that the Drude spectral weight (DSW) for *(110)*$_{240nm}$ is nearly temperature independent (Fig. 5b). But for all other samples, it increases sharply in the vicinity of $T_{IM}$ (~150K). Also, as the *(100)*$_{35nm}$/*(100)*$_{240nm}$ films contain uncondensed Drude electrons only, their DSW is 2-4 times larger than that of other films (Fig. 5a and 5b). It is generally expected that a dominant Drude contribution with increasing temperature across the *I-M* transition should diminish the CDW mode. However, we observed an increase in LSW for *(111)*$_{240nm}$ near $T_{IM}$~150K (Fig. 5d) which suggest that the background conductivity supplement the CDW mode for *(111)*$_{240nm}$ film. Whereas, the Lorentz spectral weight (LSW) for *(110)*$_{240nm}$ film keeps increasing with decreasing temperature which reasserts that more number of electrons gets condensed in the CDW phase.

To understand the carrier dynamics across various NNO films, the Drude carrier concentration ($N$) was calculated from the Drude plasma frequency ($\omega_{PD}$) using the relation, $\omega_{PD}^2 = 4\pi Ne^2/m^*$, where, $m^* \sim 6m_o$ is the effective mass and $m_o$ is the mass of the free electron (Fig 5c). An $N$ of the order of $\sim 10^{19}$ and $\sim 10^{20}$ was obtained, respectively, in the insulating and the metallic states for $(100)_{35nm}$, $(100)_{240nm}$ and $(110)_{35nm}$ films. For $(110)_{240nm}$ and $(111)_{240nm}$ films exhibiting CDW mode, the value of $N$ varied from $\sim 10^{18}$ in the insulating state to $\sim 10^{19}$ in the metallic state which are lesser than the *(100)* oriented films exhibiting Drude behaviour only. These values of $N$ are however, similar to those obtained using infrared spectroscopy $\sim 10^{18}$ whereas, Granados *et. al* estimated a higher $N$ of $\sim 10^{22}$ by assuming one electron per Ni site.[12,26] Further, the reduction in the number of electrons in the low temperature phase may be associated with the setting of the unusual AFM order whereas, the higher values of $\omega_{PD}$ above the *I-M* transition suggests weaker electronic correlations in the metallic paramagnetic phase (Fig. 5c).

**Discussions:**

Exploring the correlation between the spectral features and the structure is crucial to reveal the origin of the CDW mode in NNO.[9,8,33] The epitaxial strain is known to alter the spin and charge correlations in the NNO.[18] In present studies, the CDW mode shows a remarkable dependence on crystal orientations (Fig. 5f). The lack of CDW mode in *(100)* spectra, despite an *I-M* transition, indicates a purely SDW transition exist without a consequent CO.[33] Indeed this suggests the existence of a single Ni site (i.e. $Ni^{3+}$) for cubic/tetragonal NNO *(100)* films (Fig. 5f). This is analogous to the compressive strained $LaNiO_3$ films in which the strain, as accommodated via Ni-O bond bending and stretching, precludes the charge-ordered ground state.[33] In contrast, the CDW mode in NNO *(110)* films suggests that the orthorhombic distortion presented by *(110)* orientation induces an octahedral breathing distortion in the $Ni-O_6$ network leading to distinct Ni sites (say $Ni_1$ and $Ni_2$) in the insulating phase. As a result, the precession of bond length variation $Ni_1$-O and $Ni_2$-O may be envisaged leading to charge dissproportionation [$Ni^{3+\delta}$-$Ni^{3-\delta}$ ($\delta \sim 0.2-0.5$)] and the CO (Fig. 5f).[13] Similar scenario exists for orthorhombic NNO *(111)* films. This explanation for the CDW mode in *(110)* and *(111)* NNO films holds striking resemblance with a charge disproportionate scenario suggested for the nickelate films under tensile strain.[33]

The magnitude of CDW peak, its presence and absence for the NNO films may be controlled by the amount of orthorhombic distortion. The orthorhombic distortion is quantified in terms of orthorhombicity factor, $s = b/a$, where $a$ and $b$ are in-plane lattice constants.[34] The '$s$' is unity for cubic/tetragonal structures and greater than one for orthorhombic structures. It is clearly seen in Fig. 5(f-g) that the tetragonal *(100)* films having $s=1$ lack any CDW mode. However, this mode emerges and strengthens as '$s$' increases beyond unity in the orthorhombic *(110)* and *(111)* films; being strongest in most distorted *(110)*$_{240nm}$ film [Fig.5(f-g)]. A clear illustration for the emergence of CDW peak for NNO films while traversing from tetragonal to orthorhombic symmetry is depicted in Fig. 5(f-g). Further, as per Landau theory, two types ground state were predicted for NNO, namely, site centered and bond centered.[19-20] The former occurs in cubic/tetragonal structure, results in equal moments at all Ni sites and, hence, does not induce the CO. This proposition is consistent with lack of any CDW mode in tetragonal NNO *(100)* films. In the latter case, the bond centered SDW is driven off-center in the orthorhombic nickelates, and hence, the CO is induced as a secondary effect.[19,20] This too agrees with the presence of a CDW mode in *(110)* and *(111)* NNO films. Overall, our observations of CDW mode very well agrees with the predictions of Landau theory and underline the subtlety of electron-lattice coupling in manifestation of this phenomenon in nickelates.

Finally, we comment on the emergence of the CDW mode due to the presence of quasi-1D charge ordering in NNO. In the framework of Landau theory of NNO, some sections of Fermi surface are flat and display nesting tendencies, respectively, indicating the 1D character and enhanced susceptibilities at the certain wave-vectors leading to SDW/CDW formation.[19-20] This is analogous to the Peierel's idea of the CDW formation for a classical 1D metal.[1] A gap Δ~22 meV for NNO suggests the localization of charges carriers and the electron-phonon coupling. The Hartree Fock calculations also predict that a gap in the density of states of NNO may open between the polarized electrons in the spin-up $e_g$ band of $Ni^{2+}$ and unpolarized electrons of $Ni^{4+}$.[16,17,19,20] To relate this gap with the nesting tendency of the Fermi surface, we define a quantity Φ as the ratio of DSW and total spectral weight (TSW) (Fig. 5e).[28] The Φ is a measure of the fraction of Fermi surface not affected by the CDW state as a smaller (larger) $\Phi$ indicates the Fermi surface is largely gapped (ungapped). For *(110)$_{240nm}$* film, a $\Phi \sim 0.5$ suggest that nearly half of the Fermi surface is gapped (Fig. 5e). This imperfect nesting of the Fermi surface results from the electronic inhomogenieties in the region of coexisting metallic and insulating phases below the *I-M* transition. For *(111)$_{240nm}$* / *(110)$_{35nm}$* films, the $\Phi$ varies as ~0.3/0.6 (<$T_{IM}$~140K) and 0.6/0.9 (>$T_{IM}$), suggesting that as the temperature is lowered a larger fraction of the Fermi surface is gapped. This forms a condensed CDW mode, which broadens as Fermi surface nesting depletes above $T_{IM}$. Furthermore, an $\Phi \sim 1$ for the *(100)* oriented NNO films suggest ungapped Fermi surface, which lacks the CDW mode. Another key attribute of the CDW mode is its anisotropic character which, in present case, is expected from the directional character of the Ni $e_g$ orbital of NNO.[1,7,19,20] This was evidenced as the THz conductivity along dissimilar in-plane axes (001) and (011) of *(110)$_{240nm}$* differed by a factor of ~3 which directly corroborates with the ρ-T measurements (Supplementary information S4). Thus observed anisotropy in σ is consistent with the theoretical description of the complexity in the directionality and symmetry of the $e_g$ orbital and the flatness of the Fermi surface and may lead to the CDW formation in NNO.[19,20]

To sum up, we present the first experimental evidence of the manifestation of charge-density-waves and their structural control in the perovskite nickelates. By the means of a novel structural route to instil the collective CDW mode, it is established that the charge-ordered phase of NdNiO$_3$ is incipient upon structural modification, is of CDW type and suggests the electron-lattice coupling in weak localization limit. This low-energy mode in nickelates may serve as new channel of collective charge transport, which can redefine their quantum criticality having potential implications in the oxide electronics based on them. It is added that the charge-order condensing into CDW may be a generic feature of the ordered electronic phases in correlated perovskite oxides and can inspire CDW explorations in a wide range of similar systems.

**Method:**

**Sample Fabrication:** The NdNiO$_3$ (NNO) epitaxial thin-films were deposited on LaAlO$_3$ (LAO) substrate with two thicknesses of 240nm and 35nm along each of *(100), (110)* and *(111)* [and named as *(orientation)*$_{thickness}$, for example 240nm NNO films along (110) crystal axes as *(110)*$_{240nm}$ using pulsed laser deposition technique. The deposition was carried out using a 248nm KrF excimer laser at a repetition rate of 4Hz and a laser fluence of 1.5 J/cm$^2$. The substrate was kept at a temperature of 700ºC. An O$_2$ pressure of 35 Pa was maintained during deposition and post-deposition the films were annealed in 1 kPa of O$_2$ for 5 minutes. Thickness of the bilayers was measured using a surface profiler Dektak XT Advanced. The high resolution X-ray diffraction measurements of all the NNO films were performed on a PANalytical Empyrean. Resistivity $\rho$-T measurements were performed using a standard four probe resistivity method on a Physical Property Measurement System (PPMS) Quantum design, USA.

**Terahertz measurements:** THz-TDS data were collected on a photoconductive (LT-GaAs) antenna-based TERA K-8 spectrometer from Menlo Systems GmbH (Germany) equipped with a cryostat JANIS (SHI-4-5) cryostat. The terahertz beam was passed from both the NNO sample and LAO bare substrate (used as reference) at all the temperatures from 5K to 300K. The amplitude and phase shift of the NNO thin films was deduced using the Fast Fourier Transform (FFT) of the transmitted THz pulse. Thus obtained complex refractive index $\tilde{\varepsilon}(\omega) = \varepsilon_a(\omega) + i\varepsilon_b(\omega)$ provides the real part of optical conductivity using the relation, $\sigma(\omega) = \varepsilon_o \omega \varepsilon_b(\omega)$, $\varepsilon_o$ is the permittivity of free space.[7,8]

**Data Analysis:** The optical conductivity data was analysed using the phenomenological Drude-Lorentz (D-L) model,[28] for the optical conductivity σ-E spectra of the *(110)*$_{240nm}$, *(111)*$_{240nm}$ *(110)*$_{35nm}$ films. The standard Drude-Lorentz equation is

$$\sigma(\omega) = \frac{1}{4\pi} \frac{\Gamma_d \omega_{PD}^2}{\Gamma_d^2 + \omega^2} + \frac{1}{4\pi} \frac{S_n^2 \omega^2 \gamma_n}{(\omega_n^2 - \omega^2)^2 + \omega^2 \gamma_n^2} \qquad (D+L)$$

Where, the fitting parameters $\varepsilon_0$ is permittivity of free space, $\omega_{PD}$ is the Drude plasma frequency, $\Gamma_d$ is Drude scattering rate, $\omega_n$ is center peak frequency, $\gamma_n$ is the width and $S_n^2$ is the mode strength for the $n^{th}$ Lorentz harmonic oscillator.

**Acknowledgments**

This work is supported by the Department of Science and Technology, New Delhi under the research project SR/S2/LOP-13/2010.

**Author contributions**

R.R. and D.S.R. conceived and designed the experiments. R.R. and P.P. carried out the experiments and analysed the data. R.R. and D.S.R. wrote the paper. R.R., P.P., S.S.P. and D.S.R. discussed the results and commented on the manuscript.

Figure 1:

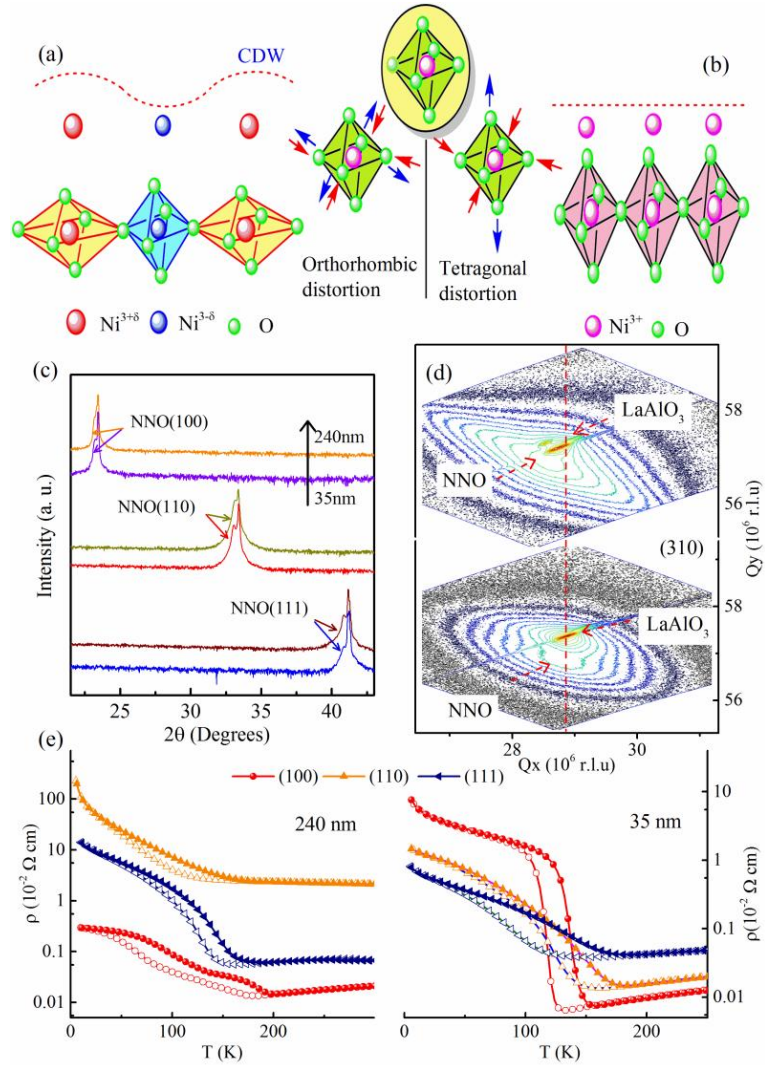

Figure 1: (a) Shows the orthorhombic distortion and (b) depicts the tetragonal distortion of the oxygen octahedral centred around Ni transition metal oxide ion in the NNO, (c) depicts the *θ-2θ* scan of the NNO thin-films on LaAlO$_3$ *(100)*, *(110)* and *(111)* substrate, (d) shows the reciprocal space maps (RSM) of NNO along asymmetric peak reflection of (310) in the upper panel for *(110)*$_{240nm}$ and in lower panel for *(110)*$_{35nm}$ film and (e) depicts the resistivity (ρ) versus temperature (T) behaviour of various NNO films, where solid and open symbols depicts the ρ-T measurements while cooling and heating runs.

Figure 2:

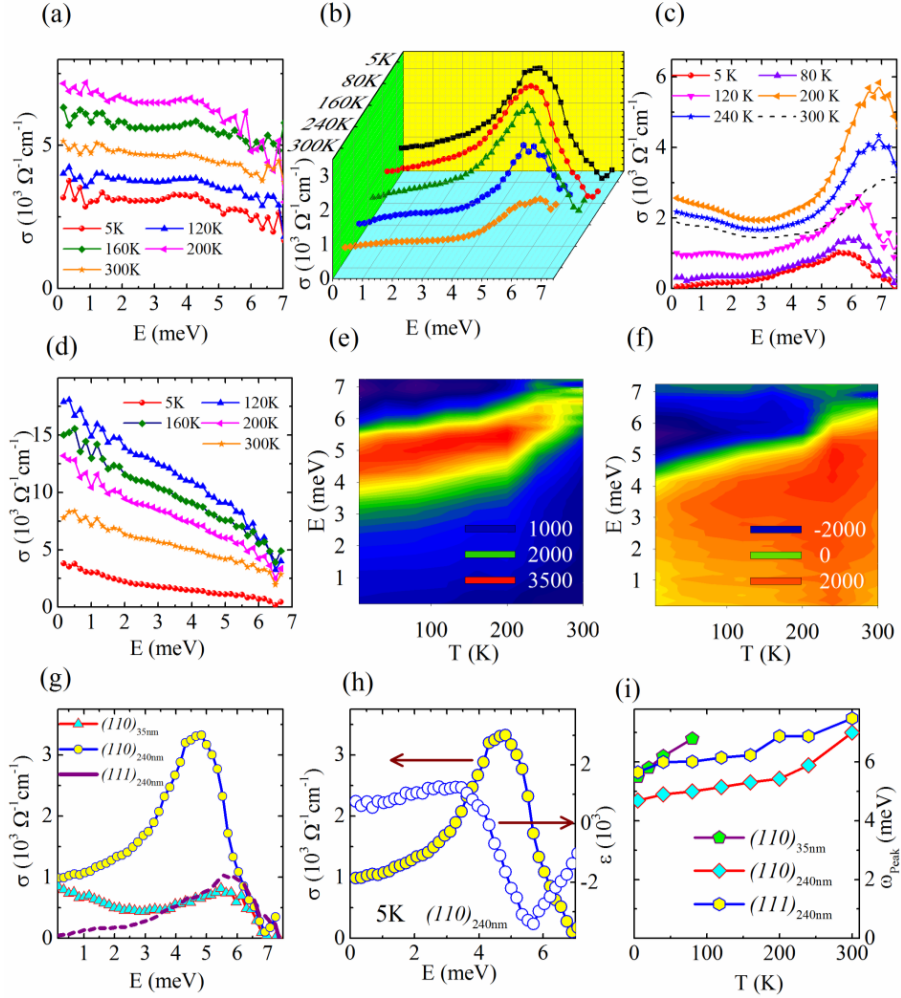

Figure 2: Shows the optical conductivity (σ) versus terahertz photon energy (E) for (a) of *(100)*$_{240nm}$ ,(b) of *(110)*$_{240nm}$ ,(c) of *(111)*$_{240nm}$ ,(d) of *(100)*$_{35nm}$ NNO films (e and f) shows the temperature dependence of σ (where red region shows the CDW peak shift with increasing temperature) and dielectric constant (ε) for *(110)*$_{240nm}$ ,(g) shows the σ-E spectra of *(110)*$_{240nm}$, *(111)*$_{240nm}$ and *(110)*$_{35nm}$ NNO films at a temperature of 5K ,(h) shows the dielectric dispersion of σ-E spectra of *(110)*$_{240nm}$ film (i) shows the CDW peak shift of the *(110)*$_{240nm}$, *(111)*$_{240nm}$ and *(110)*$_{35nm}$ NNO films

Figure 3:

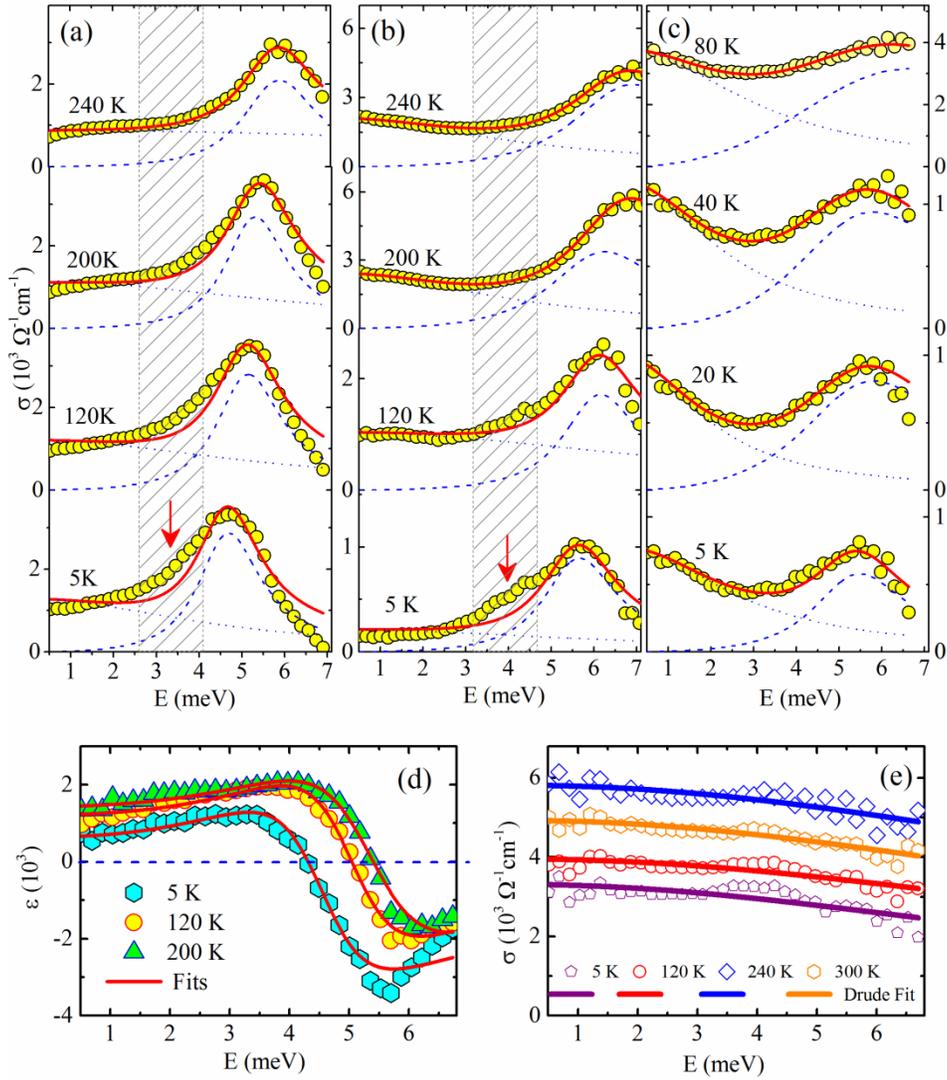

Figure 3: (a, b and c) shows the Drude-Lorentz (D-L) fits for $(110)_{240nm}$ ,$(111)_{240nm}$ and $(110)_{35nm}$ for NNO films, here symbols, solid line, dashed line and dotted lines depict the experimental data, D-L fit, Lorentz contribution and Drude contribution. (d) Shows the Dielectric constant (ε) at various temperature for $(110)_{240nm}$ and solid line depicts fits as per $\varepsilon(\omega) = \varepsilon_\infty + \omega_{PL}^2 (\omega_{peak}^2 - \omega^2)/(\omega_{peak}^2 - \omega^2)^2 + \omega^2 \Gamma^2$ and (e) shows the Drude fits (solid line) at various temperature for $(100)_{240nm}$ films.

Figure 4:

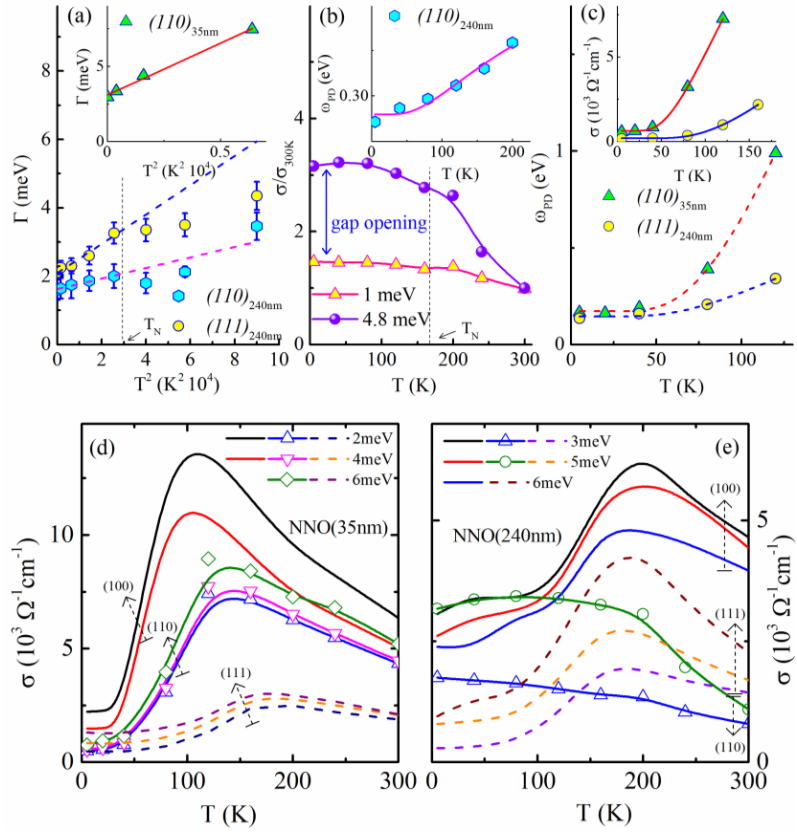

Figure 4: (a) Shows the scattering rate (Γ) versus square of the temperature (T) for *(110)$_{240nm}$* and *(110)$_{240nm}$* and *(110)$_{35nm}$* (inset a) [where symbols denote experimental data and dotted and solid line lines depict the fit as per equation $\Gamma(T) = \Gamma_o + AT^2$, (b) shows the temperature variation of the normalized conductivity for *(110)$_{240nm}$* film ,(c) depicts the optical conductivity at (0.8 meV) versus temperature for *(111)$_{240nm}$* and *(110)$_{35nm}$* films inset (b) and (c) shows the Drude plasma frequency versus ($\omega_{PD}$) versus temperature for *(110)$_{240nm}$* and *(111)$_{240nm}$* and *(110)$_{35nm}$*, the symbols show the experimental data while solid line/dotted line depicts the fits as per equation $\sigma(T)[\omega(T)] = A_o \exp[-\Delta/k_\beta T] + B$, and (d & e) shows the temperature dependence of optical conductivity (σ) at different photon energy for NNO films.

Figure 5:

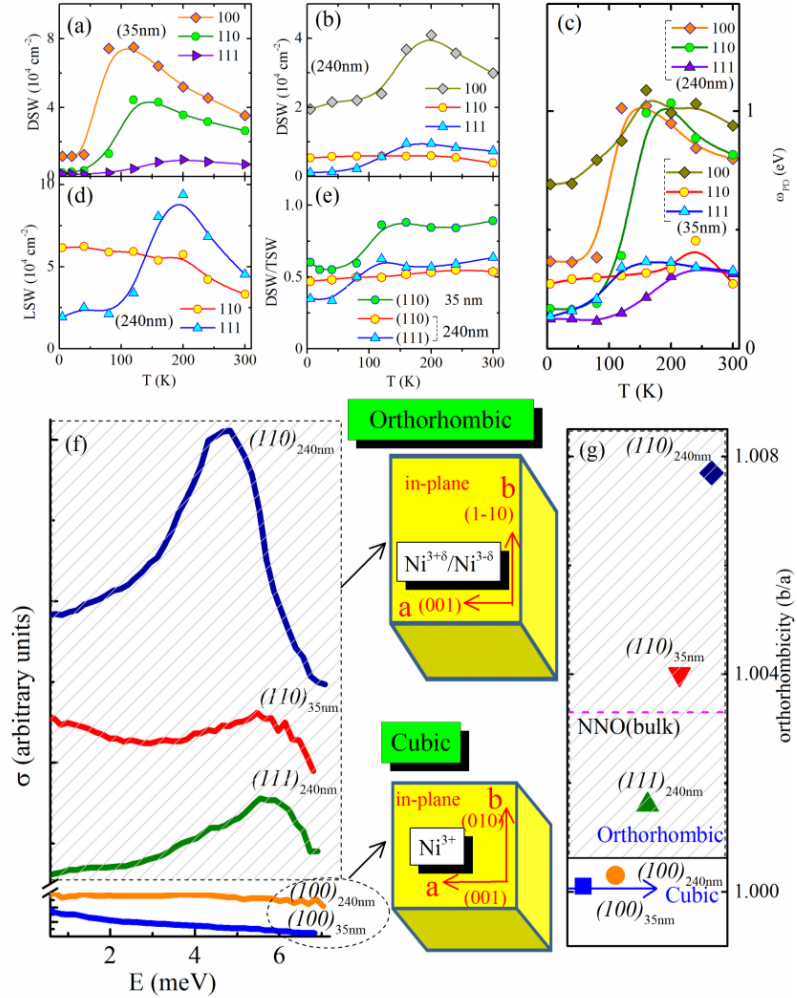

Figure 5: Temperature dependence of the experimental Drude spectral weight (DSW) for (a) 35nm NNO film and (b) 240nm NNO films, (c) is the temperature variation of the Drude Plasma frequency ($\omega_{PD}$) for NNO films, (d) Temperature variation of the experimental Lorentz spectral weight (LSW) for 240nm *(110)*, *(111)* NNO films, (e) shows the ratio of DSW/Total spectral weight(TSW) for NNO films and (f-g) shows the emergence of CDW peak in NNO films for dissimilar orientations with increasing orthorhombicity (*s*=b/a) which is the ratio of two orthogonal in-plane lattice constants *a* and *b*.